\documentclass[prb,aps,twocolumn]{revtex4}
\usepackage{epsf}
\usepackage{epsfig}
\usepackage{graphics}
\begin{document}

\title{Novel effects of localization due to `intrinsic disorder' in the `two-fluid' model for manganites}

\author{Prabuddha Sanyal$^{1,2}$}
\author{V.B. Shenoy$^{2}$}
\author{\mbox{H. R. Krishnamurthy$^{2,3}$}}
\author{T. V. Ramakrishnan$^{2,3,4}$}

\affiliation{$^{1}$ School of Physics, Hyderabad Central University, Hyderabad- 500046}
\affiliation{$^{2}$Centre for Condensed Matter Theory (CCMT),
Department of Physics, Indian Institute of Science, Bangalore
560012, India.}
\affiliation{$^{3}$Jawaharlal
Nehru Centre for Advanced Scientific Research, Bangalore 560064,
India.}
\affiliation{$^{4}$Department of Physics, Banaras Hindu
University, Varanasi 221005, India.}

\begin{abstract}
We discuss the effects of a novel polaronic disorder in the recently proposed  
two-fluid model for manganites. Using effective field theory as well as direct numerical
 simulations, we show that this disorder can have dramatic effects in terms of the 
 transition from ferromagnetic insulator to ferromagnetic metal upon hole-doping, including
 an Anderson localized regime where variable range hopping may be observed.
\end{abstract}

\maketitle

\section{Introduction}

  Doped rare-earth manganites A$_{1-x}$AE$_{x}$BO$_{3}$ (A=Rare-Earth, AE=Alkaline Earth, B=Mn) are transition metal oxides where there is an intricate interplay of charge, orbital, spin and phonon
  degrees of freedom. No less important, however, is the role of disorder, which has dramatic
 effects upon the charge ordering, magnetic transition, etc~\cite{Tokurarev}.
  While many different types of disorder has been discussed in the context of
  manganites (eg. A-site disorder, B-site disorder etc.)~\cite{Asitedis,Al-sub,Cr-sub},
  in this paper we will introduce a
  novel kind of disorder apparently present in manganites but not widely talked about,
  and demonstrate its effects using the two-fluid ('$\ell$-$b$') model Hamiltonian~\cite{Hassan}
  which was 
 proposed earlier for manganites.
 
The basic physics of manganites involve a coexistence of fast moving band electrons, and self-trapped JT polarons, interacting by Coulomb interaction, and strongly coupled to a background of core spins.
 The `$\ell$-$b$' model is an
effective low energy Hamiltonian which implicitly captures the
crucial effects of these interactions and the quantum dynamics of
the JT phonons. It invokes two types of $e_g$ electrons, one {\em
polaronic} and {\em localized} ($\ell$), and the other {\em
band-like} and {\em mobile} ($b$), and is given by,

$$H_{\ell b}  =  (-E_{JT}-\mu)\sum_{i,\sigma} n_{\ell i\sigma} -\mu
\sum_{i,\sigma}n_{bi\sigma} $$
$$+ U_{dd} \sum_{i, \sigma}n_{\ell
i\sigma}n_{bi \sigma}  -  t\sum_{<ij>, \sigma}(b^{\dagger}_{i,
\sigma} b_{j,\sigma } + H.C.) $$
\begin{equation}
 - J_{H}\sum_{i}({\vec {\sigma}_{\ell
i}} + {\vec {\sigma}_{bi}}) \cdot {\vec{S}}_{i} -
J_{F}\sum_{<ij>}{\vec{S}_{i}} \cdot {\vec{S}_{j}}
\label{lbham}
\end{equation}
The  {\em polaronically trapped }  `$\ell$' species has site
energy $-E_{JT}$ ($\sim$ 0.5 $eV$), and an exponentially reduced hopping ($\sim$ 1
$meV$) which has been neglected, while the {\em non-polaronic}
`$b$' species (site energy $0$) has undiminished hopping $t \sim$
0.2 $eV$. $U_{dd}$ ($\sim$ 5 $ eV$) is the effective on-site
Coulomb repulsion between the `$\ell$' and `$b$' species and
$J_{H}$ the aforementioned Hund's coupling. $J_{F}$ is a novel
ferromagnetic {\em virtual double-exchange} coupling ($\sim 10~
meV$) between the core spins, which arises from {\em{virtual, fast
(adiabatic)}} hopping processes of the `$\ell$' electrons to
neighboring sites and back, leaving the local lattice distortion
unrelaxed~\cite{Hassan}. The chemical potential, $\mu$, imposes
the doping determined filling constraint: $\sum_{\sigma}(\langle
n_{\ell \sigma}\rangle + \langle n_{b\sigma}\rangle) = (1-x)$.
At T=0, in the fully polarized ferromagnetic phase, it reduces to a form
very similar to the Falicov-Kimball model~\cite{fkm-rev}.

 While disorder is commonly grouped into `annealed' and `quenched',
 we make a further distinction in the context of this $\ell-b$ model for manganites,
 namely: \emph{intrinsic} and \emph{extrinsic}.
  We note that from the basic premises of the 
 $\ell-b$ Hamiltonian, the $\ell$ electrons are polaronically trapped at specific lattice sites, and have no
  quantum dynamics. They however, equilibriate according to classical statistical mechanics. The mobile $b$
 electrons hop in the background provided by these static $\ell$ polarons. Hence, the $\ell$ species represents
 an annealed disordered background to the mobile $b$ species. We refer to this disorder as `intrinsic', since
 it is `self-generated' by the system under appropriate conditions. 
 On the other hand, disorder due to doping at the
 A-site by alkaline
 earth atoms, or doping at the B-site by Aluminium, Chromium etc., is referred to
 as `extrinsic'. Thus, the $b$ electrons face two kinds of disorder of the compositional type:
 namely \emph{intrinsic, annealed} disorder due to the $\ell$ polarons~\cite{TVRbook},
 and \emph{extrinsic, quenched} disorder due to foreign dopant atoms. The effects of quenched
 disorder of the extrinsic type has already been considered 
 in some detail in the literature, both at the abstract level~\cite{MackinnonRev,LeeRamaRMP},
 and in the specific context of manganites~\cite{Pinaki1,Sanjeev}.
 In this paper we discuss the effects of this novel, intrinsic
 disorder in the context of the $\ell-b$ Hamiltonian for manganites.
   
 The $\ell-b$ model has been studied using Dynamical Mean Field Theory (DMFT)
 in Ref~\cite{Hassan}, which is exact in the limit of large dimensions. 
 Due to the specific symmetries of the $\ell-b$ model, discussed
 in Ref~\cite{Hassan}, it becomes exactly solvable in this limit, enabling a complete study
 of the phase diagram. However, DMFT, like its counterpart 
 CPA (Coherent Potential Approximation), does not include effects of Anderson localization
 which is a crucial effect of disorder at low temperatures,
  and which we want to focus on in this paper. 
 The reason for this is discussed in more detail below. Hence, in this paper, we study
 the $\ell-b$ model using an approach pioneered by Vollhardt and Economou {\it et al}
 which can take us beyond the DMFT approximation. We also compare our results with studies
 of Inverse Participation Ratio (IPR) done using a direct numerical simulation.

\section{DMFT of l-b model: summary}

 In the $\ell-b$ Hamiltonian, the mobile $b$ electrons move in an annealed disordered 
 background of site-trapped $\ell$ polarons. The DMFT approximation can be used to 
 study the electronic and transport properties of this Hamiltonian.
 We first solve the DMFT
 exactly using semicircular DOS, and in the limit $U_{dd}->\infty$~\cite{Hassan}, in the fully spin-polarised
 ferromagnetic phase at T=0~\cite{cubic}. The DMFT approximation reduces the complicated
many-site problem given by Eqn~\ref{lbham} into an effective single-site problem, represented
by the action: 
$${\mathcal A}_{DMFT}(n_{\ell})=\sum_{n}[({\mathcal G}_{bath}^{-1}(i\omega_{n},\Omega_{z})-U_{dd}n_{\ell})\overline{b}_{n}b_{n}]$$
\begin{equation}
-(E_{JT}-\mu)n_{\ell}
\label{Eq: action_DMFT}
\end{equation}
where $n_{\ell}$ is the fractional occupancy of $\ell$-electrons at the DMFT site, and ${\mathcal G}(\omega)$ is the Weiss or bath Green's function. Solving this single site `impurity' problem, we obtain the full impurity Green's function $G(\omega)$. Since this action is similar to that
for the Falicov Kimball model~\cite{fkm-rev}, hence it can be solved exactly.
  We can also obtain the
 exact self-energy for the b-electrons using the Dyson's equation:
\begin{equation}
\Sigma(\omega)=\mathcal G^{-1}(\omega)-G^{-1}(\omega) 
\end{equation}
This expression is given by:
\begin{equation}
\Sigma(\omega)=\left(1-\frac{1}{W_{0}}\right){\mathcal G}^{-1}(\omega)
\end{equation}
where
\begin{equation}
{\mathcal G}(\omega)=\frac{2}{\omega+\sqrt{\omega^{2}-W_{0}D^{2}}}
\label{bath_greens_func}
\end{equation}
and $W_{0}$ is the probability of there being no $\ell$-polaron on the impurity site of the
 DMFT, to be determined self-consistently. For low dopings, $W_{0}\approx x$. $D$ is the
half-bandwidth of the bare semicircular DOS.

 The optical conductivity of a thermodynamic system can be obtained within linear response, using the Kubo
 formula:
\begin{equation}
{\mathcal R}\overline{\sigma}(\omega^{+},q\rightarrow0)=-\frac{1}{\omega d}\sum_{\alpha=1}^{d}{\mathcal I}\Pi_{\alpha\alpha}(\omega^{+},q\rightarrow0)
\label{eq: Kubo_formula}
\end{equation}
where $\Pi_{\alpha\beta}(\omega,q)$ is the well-known current-current correlation function.
In the DMFT limit of large dimensions, only the elementary particle-hole Drude 
bubble for this correlation
 function survives, while all higher terms involving the irreducible vertex become negligible
~\cite{GeorgesRMP}. The contribution to the optical conductivity 
for a hypercubic lattice then
becomes~\cite{GeorgesRMP}:
\begin{equation}
\sigma(i\omega_{n})=\frac{1}{\omega\beta}\sum_{k\nu_{n}\sigma}\frac{1}{d}\sum_{l=1}^{d}
4sin^{2}(k_{l})G(\vec{k},i\nu_{n})G(\vec{k},i\nu_{n}+i\omega_{n})
\label{conduct_DMFT}
\end{equation}

\section{Beyond DMFT: Self-consistent theory of localization} 

  The DMFT approximation does not
 capture the physics of \emph{Anderson Localization}, which may well be important for the low temperature
 transport behaviour of manganites having both intrinsic and extrinsic disorder. In particular, the 
  Metal Insulator Transition (MIT) as a function of doping at low temperature is expected to be strongly
 influenced by localization physics. To investigate this aspect, we go beyond DMFT and study the mobility
 edge behaviour as a function of doping using the \emph{Self-consistent Theory of Localization} (STS)
 developed by Vollhardt {\it et. al.} \cite{Vollhardt}. Economou {\it et. al.} 
 had proposed a particularly simple, and practically
 useful prescription for finding the mobility edge by mapping the STS equations formally to
 equations for bound state formation in a potential well~\cite{Economoubrief}.
 We use Economou's approach in this paper, which
 we refer to hereafter as \emph{Potential Well Analogy} (PWA).  
 We first find the single particle Green's function
 for the $b$-electrons, as well as the self-energy, using the DMFT approximation at zero temperature.
 The dc conductivity
 of the system is also found within the DMFT approach. Then we plug these quantities into the PWA equations
 to obtain the mobility edge at any particular doping.

 In order to incorporate
 localization physics, one has to add correction terms to the Drude conductivity, obtained from 
  `maximally crossed' diagrams in the theory of weak localization~\cite{Bergmann}:  
\begin{equation}
\sigma(\omega)\approx\sigma_{0}-\frac{2e^{2}}{\pi\hbar(2\pi)^{d}}\int^{1/l_{el}}_{1/l_{in}} \frac{d\vec{k}}{k^{2}-\frac{i\omega}{D_{0}}}
\label{loc_corr_sigma}
\end{equation}
 where $\sigma_{0}$ is the Drude conductivity, $l_{el}$ is the elastic mean free path, while $l_{in}$
 is the inelastic mean free path. Using the Einstein relation between conductivity and diffusivity,
\begin{equation}
\sigma=2e^{2}\rho D_{0}
\label{Einstein}
\end{equation}
 one can rewrite eq~\ref{loc_corr_sigma} purely in terms of diffusivites. 
Now, the theory was made self-consistent by Vollhardt and Wolfle \cite{Vollhardt} by replacing $D_{0}$ by the full 
 $D(\omega)$ on the denominator of the integrand.

\begin{equation}
 D(\omega)= D_{0}-\frac{1}{\rho\pi\hbar(2\pi)^{d}}\int^{1/l_{el}}_{1/l_{in}} \frac{d\vec{k}}{k^{2}-\frac{i\omega}{D(\omega)}}
\label{STS}
\end{equation}

The localization
 length $\lambda$ was found by identifying $\frac{-i\omega}{D(\omega)}\rightarrow \frac{1}{\lambda^{2}}$
 in the insulating regime.

Economou and Soukoulis \cite{Economoubrief} pointed out that this equation has the same formal structure
 as the equation which determines the decay length of a bound state in a potential well.

\begin{eqnarray}
\frac{1}{\Omega|V_{0}|}=\frac{2m^{*}}{(2\pi)^{d}\hbar^{2}}\int \frac{d^{d}k}{k^{2}+k_{b}^{2}}
\label{PWA_TB}
\end{eqnarray}
where $\Omega$ is the volume of the primitive unit cell, $k_{b}$ is the inverse decay length
of the bound state, while $|V_{0}|$ is the well depth, and $m^{*}$ is the effective mass
near the band bottom.

For the case of tight binding on a hypercubic lattice, one particular 
site has the well potential $V_{0}$. Since the integral is over the Brillouin zone,
one establishes a connection between the lattice constant $a$ and the mean free path $l_{el}$,
neglecting $1/l_{in}$ at low $T$. Now, knowing the minimum depth of the potential well necessary
for the appearance of a bound state\cite{Economoubook,Economoubig}, one can estimate
 the mobility edge.

The mean free path $l(E)$ is found as usual:
\begin{eqnarray}
l(E)&=&v(E)\tau(E) \\
    &=&\frac{v_{0}(E-{\mathcal R}\Sigma(E))\hbar}{2{\mathcal I}\Sigma(E)}
\label{l_v_tau}
\end{eqnarray}
where $v_{0}(E)$ is defined as the average speed of Bloch waves over the surface of constant energy $E$.

 The Drude conductivity $\sigma_{0}(E)$ can be found either in its full glory
  from Eq~\ref{conduct_DMFT}
or using the weak scattering approximation:
\begin{equation}
\sigma_{0}(E)\approx \frac{2e^{2}}{(2\pi)^{d}d\hbar}S_{0}(E-{\mathcal R}\Sigma(E))l(E)
\label{weakscat_mobedge_crit}
\end{equation}
 where $S_{0}(E)$ is the area of the surface of constant energy.

\section{Results}
 As discussed in detail in \cite{Hassan},
 the $\ell-b$ model shows a transition from Ferromagnetic Insulator to Ferromagnetic metal,
 as the doping is increased, just as observed in real manganites. This is because
 of depletion in the number of trapped Jahn-Teller polarons, and 
 correponding decrease in scattering and increase in paths available for
  hopping, upon hole-doping. Within DMFT, for $U\rightarrow\infty$,
  the critical doping for this Insulator-Metal
 transition (IMT) can be obtained analytically as 
\begin{equation}
  x_{c}^{DMFT}=\left(\frac{E_{JT}}{D}\right)^{2}
\label{x_crit_DMFT}
\end{equation}
which corresponds to the doping where the bottom edge of the $b$-band
hits the sharp level at energy $E_{JT}$ correponding to Jahn-Teller trapped
$\ell$-polarons. Since for $U\rightarrow\infty$ the DMFT $b$-electron DOS is proportional
to the imaginary part of the bath Green's function mentioned above, they have the
same bandwidth $\tilde{D}=\sqrt{W_{0}}D$, as is evident from Eq~\ref{bath_greens_func}.
Hence, when the $b$-band bottom hits the polaronic level, $\tilde{D}=E_{JT}$, giving
the relation Eq~\ref{x_crit_DMFT} above.
For $x<x_{c}$, the entire $(1-x)$ fraction of electrons were polaronically
trapped in the local $E_{JT}$ level, giving a polaronic insulator. For $x>x_{c}$, the
$b$-band states are available for the electrons to delocalize, which according to DMFT,
should immediately lead to a metallic state. We would like to improve upon this picture
in this paper by bringing in the effect of Anderson localization.

  We calculate the relaxation time $\tau$, mean free path $l$
 and transport DOS $v(E)S(E)$ for a cubic lattice and a single-orbital, tightbinding band. 
 The DMFT parameters are chosen so that the bottom edge of the $b$-band 
 hits the sharp $\ell$-level at
 a doping ($x_{c}^{DMFT}$) of about $0.25$. On account of the exact analytic relation
 mentioned above, this is accomplished by choosing, {\it eg}.,
  $E_{JT}$ to be $-3$ in units of hopping.
 This is because, for a cubic lattice, the half bandwidth is $6t$, which we identify 
 with the half-bandwidth $D$
 of the semicircular DOS mentioned above. Thus $D=6$ in units of hopping $t$.
 The mobility edges are calculated 
  both using the full conductivity, and the weak-scattering assumption as in 
 Eq~\ref{weakscat_mobedge_crit}.

\begin{figure}[htb]
\epsfysize=6cm
\epsfxsize=8cm
\centerline{\epsfbox{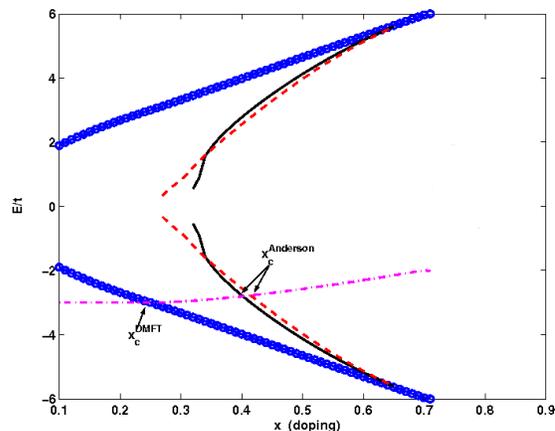}}
\vspace*{-3cm}
\vspace{1 in}\caption{(colour online) mobility edge trajectory vs doping. Circles (blue): band
edges. Dash-dot line (red): Fermi energy. Solid Lines (black): Mobility edges using actual DMFT conductivity. Dashed lines (red): Mobility edges using weak scattering approximation. Parameters: $E_{JT}/t=-3,D/t=6$}
\label{fig: mob_edge_DMFT_STS}
\end{figure}

 The mobility edge trajectories obtained by both methods are plotted as a function of the
 doping in Fig~\ref{fig: mob_edge_DMFT_STS}. One finds that for either method,
 there are two mobility edges occuring symmetrically
 about the centre of the band, between which lie the extended states. States lying between any one mobility
 edge and the band edge on the same side are localized, since it is the band tails which get localized first.
 For low doping values 
 below $x_{c}^{DMFT}$, 
 all states in the band are localized, and both the mobility edges are coincident at the band centre.
 As the doping increases, both the mobility edges proceed outwards from the band centre. 
 Simultaneously, the bandwidth also increases (roughly as $\sqrt{x}D$ for low doping values),
 due to reduced scattering from polarons.
The rate at which
 the mobility edges move apart in energy with increasing doping is greater than the rate at which the 
 bandwidth increases, so that ultimately, the mobility edges meet the corresponding band edges at a
 large enough doping ($\approx 0.6$ for this parameter set). This means that beyond this doping value all
 states in the band are extended. This makes sense, because nearabout this large doping, there are no more
 $\ell$-polarons left ($W_{1}=0$). This means that the $b$-electrons do not encounter any scattering, and
 the bandwidth is at its maximum. It is to be noticed that the mobility edge trajectories calculated by
 the two methods differ significantly only for small doping, while they coincide for large doping. This is
 to be expected, since for smaller doping values, there are a substantial number of polaron scatterers, so that
 the weak-scattering assumption is not justified.
 
\begin{figure}[hbt]
\epsfysize=6cm
\epsfxsize=8cm
\centerline{\epsfbox{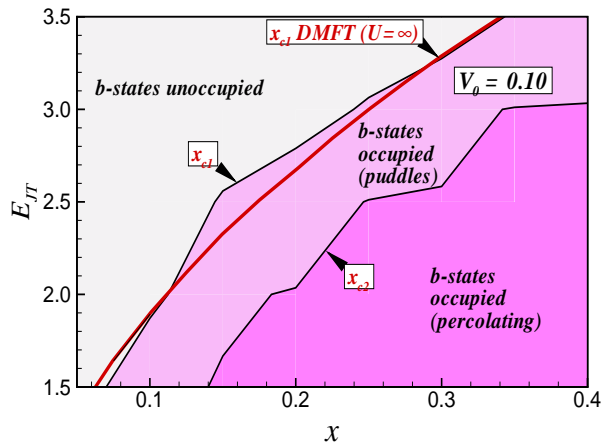}}
\vspace*{-3cm}
\vspace{1 in}\caption{$E_{JT}$ vs $x$ showing  various phase boundaries, along with the DMFT parabola for comparison.}
\label{fig: Ejt_x_shenoy}
\end{figure}

 We have also plotted the
 self-consistently determined value of chemical potential from the DMFT, on the same graph. It is found that
 even though the $b$-band states begin to become occupied for $x>x_{c}^{DMFT}$, but for a range of x thereafter,
 the occupied states are localized. The chemical potential proceeds towards the band centre, and meets the
 lower mobility edge at a doping $x_{c}^{Anderson}$ of the order of $0.4$. Hence, for intermediate values
 of doping $x_{c}^{DMFT}<x<x_{c}^{Anderson}$, the system is an \emph{Anderson Insulator}, although DMFT predicts
 that it should be metallic.

\begin{figure}[hbt]
\epsfysize=6cm
\epsfxsize=8cm
\centerline{\epsfbox{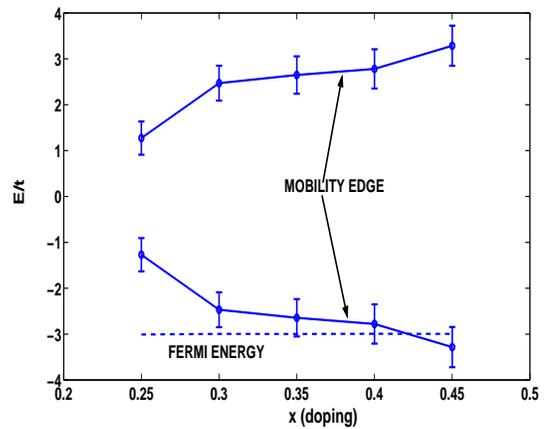}}
\vspace*{-3cm}
\vspace{1 in}\caption{mobility edge trajectory vs doping from real space numerical simulation}
\label{fig: mob_edge_RSNS}
\end{figure}

 We now compare this result, obtained using effective field theories, 
 with a more realistic, real space, finite-size simulation of the $\ell-b$ model (in the fully polarised 
 ferromagnetic phase at T=0), done by V. Shenoy {\it et.al.}~\cite{Shenoy}. Details of this
 simulation will be found in Ref~\cite{Shenoycondmat}.
 In this simulation, in addition to the Hamiltonian described before, 
 there is a long range Coulomb interaction between
 charges. This long-range Coulomb interaction (treated using Hartree-Fock approximation)
  prevents phase separation, 
 and bring about a somewhat `homogeneous' mixture of $\ell$ and $b$, as is assumed in DMFT. The presence of this
 long range Coulomb interaction, along with the compensating negative charge for the dopant atoms, gives an 
 additional source of disorder that was not included in the effective medium calculations. 
 However, this disorder is correlated,
 as compared to the uncorrelated disorder due to the $\ell$-polarons. One finds that this 
 additional correlated disorder seems to make very little difference from DMFT results with regard to the
 `critical' doping required for electrons to start occupying band states. This is observed
 in Fig~\ref{fig: Ejt_x_shenoy}, where the phase boundary for $b$-state occupancy nearly
 coincides with the DMFT parabola given by Eq~\ref{x_crit_DMFT}.

\begin{figure}[hbt]
\vspace{0.25 in}
\epsfysize=6cm
\epsfxsize=7cm
\centerline{\epsfbox{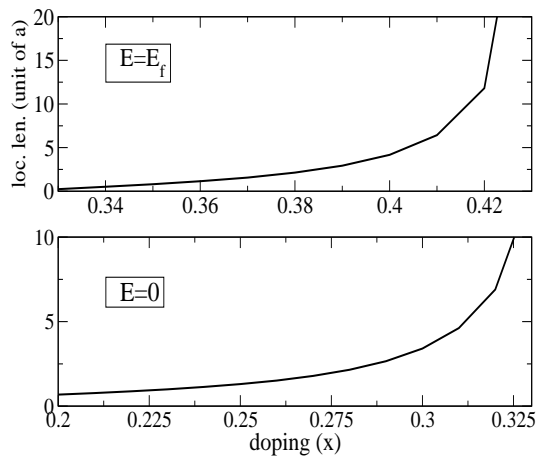}}
\vspace*{0.25in}
\caption{Doping variation of localization length evaluated at the Fermi energy (top) and at the center of the band (bottom)}
\label{fig: loclength_x}
\end{figure}

 We refer to this `critical' value of
 doping as $x_{c}^{occupancy}$. One finds that this  $x_{c}^{occupancy} \approx x_{c}^{DMFT}$. 
 However, in this
 real space simulation, even after $b$-band states begin to get occupied, the hole clumps within which
 $b$-electrons are allowed to hop (for on-site $U_{dd}\rightarrow \infty$) may not be percolating. We call 
 the `critical' doping for this percolation threshold as $x_{c}^{percolation}$, 
 and this is higher than
 $x_{c}^{occupancy}$. Moreover, even within these percolating clumps, some of the $b$-electron states may
 be Anderson localized. To check the character of these band states, we find
 the \emph{Inverse Participation Ratio} (IPR), defined as follows:
\begin{equation}
IPR=\frac{\sum_{i}^{N}|a_{i}|^{4}}{(\sum_{i}^{N}|a_{i}|^{2})^{2}}
\label{IPR_def}
\end{equation}
where $a_{i}$ is the amplitude of the state at the i-th site. For an extended state, $a_{i}$ goes as 
 $e^{i\vec{k}\cdot\vec{R_{i}}}$, so that $|a_{i}|=1$, and $IPR\rightarrow \frac{1}{N}\rightarrow 0$ for large N.
 But for a state localized
 at the j-th site, $a_{i}=\delta_{ij}$, and $IPR\rightarrow1$. Hence, IPR gives
 a good criterion for distinguishing extended states from localized ones. 

 Since different realizations of the compensating dopant charge distribution give different set of states, 
 one gets a distribution  of IPR within each small energy range inside the band. Hence, we discretize the band
 into energy intervals, and find the probability distribution for IPR within each such interval,
 by plotting histograms and corresponding frequency polygons, as discussed in detail in Ref~\cite{prabsthesis}.
  One finds from 
 Fig~\ref{fig: mob_edge_RSNS} that the mobility
 edge trajectories once again proceed away from the band centre, and towards the edges as doping increases.
 Moreover, the chemical potential, determined from the band-filling in the same simulation, lies between
 the lower band edge and the lower mobility edge for small doping. Hence, this simulation also predicts
 the existence of an Anderson Insulator phase for a certain region of doping. Moreover, the value of
 $x_{c}^{Anderson}\approx 0.4$ obtained from Fig~\ref{fig: mob_edge_RSNS}, also agrees closely with the
 effective field theory calculations described before. Thus, for $E_{JT}=3$, the
 $x_{c}^{occupancy}\approx0.25$, $x_{c}^{percolation}\approx0.33$, while $x_{c}^{Anderson}\approx 0.4$~\cite{fnote}.  

We have also calculated the localization length $\lambda$ as a function of doping using the
PAW formalism. It turns out to be
surprisingly small near $x_{c}^{DMFT}$: namely of the order of a lattice spacing, even less 
for lower doping values. As one approaches $x_{c}^{Anderson}$, $\lambda$ at Fermi energy diverges, as expected. $\lambda$ for states at the center of the band diverge for lesser $x$, since
the mobility edge trajectories converge to the center at around $x=0.3$. Variable range
hopping is therefore expected to be observed in the Anderson Insulator regime 
from $x_{c}^{DMFT}$ to $x_{c}^{Anderson}$. The resistivity
 is therefore expected to
go as~\cite{Mottbook,MottThalf} $exp(T_{0}/T)^{1/4}$, where the temperature 
 parameter $T_{0}$ is given by $T_{0}=\frac{18 \alpha^{3}}{k_{B}N(E_{F})}$, $\alpha$ being the inverse localization length.
 From our calculation of localization length,
we estimate this parameter to lie between $10^{7}-10^{8}K$ in this doping regime.

 In conclusion, we have extended the DMFT calculations for the IMT within the two-fluid model~\cite{Hassan} for manganites to incorporate localization effects. We find the existence of a
 prominent Anderson Insulator phase in addition to the polaronic, percolative insulator phases
 seen before.
 Such an Anderson Insulator phase may well explain
 the observations of variable range hopping (VRH)~\cite{LeeRamaRMP} 
 in manganites reported by experimentalists~\cite{Coeyreview,CoeyVRH,Coey2,Coey3,AKRAPL,Sun}.
 The exceptionally small $\lambda$ in the Anderson Insulating regime should translate
to a large value for the temperature parameter for the VRH, as suggested by these studies.
 In the real space simulations by Shenoy {\it et.al.}~\cite{Shenoycondmat}, 
 there exists a Coulomb glass
 phase with a soft Coulomb gap at low doping values prior to the occupancy 
 of the $b$-band states. This is due to the site-trapped $\ell$ polarons, which also
 interact with each other by Coulomb interaction.
 This should give rise to a Shklovskii-Efros (SE-VRH) $T^{1/2}$ transport~\cite{MottThalf}, which in this scenario 
  will crossover into the 
 Mott-VRH $T^{1/4}$ regime as the doping is increased, and the $b$-band states begin to get occupied, before the final Insulator-Metal transition. Interestingly, both the SE-VRH and Mott-VRH
 regimes have 
 been noticed and reported in the experimental literature~\cite{AKR_ES,SE_VRH}, and there exists
considerable debate between which one is more applicable to manganites. This is partly
 because of the technical difficulty associated with distinguishing between 
 various power-laws in the
exponent by fitting the resistivity data within a limited range of temperature~\cite{SE_VRH}. 
 Our calculation, together with the earlier ones by Shenoy {\it et.al.}, on the other
hand, postulate the existence of {\em both} the SE-VRH and Mott-VRH regimes at different values
 of doping within the phase diagram for manganites.


\begin{thebibliography}{99}
\bibitem{cubic} The self-energy, critical doping etc. for the DMFT done explicitly 
 considering the DOS for a cubic lattice, and with finite but large $U$, does not
 vary appreciably from that for semicircular DOS, and $U\rightarrow\infty$~\cite{Hassan}.
 The latter,
 however, gives the advantage of obtaining exact analytical expressions. We have repeated the
 mobility edge calculation with the DMFT done numerically using cubic DOS,
 and the results turn out be similar in all aspects.
\bibitem{fnote} It is to be noted that these values for the critical dopings themselves are not
to be considered sacrosanctly, but only as exhibiting the trend that we wish to highlight.
In order to match with real manganites, the parameters like bandwidth, $E_{JT}$ etc. has to
  be chosen appropriately so that the various $x_{c}$ have more realistic values.
\end{thebibliography}
\end{document}